\documentclass[12pt]{iopart}
\usepackage{epsfig}
\begin{document}

\title[]{$J/\psi$ production and nuclear effects for d+Au and p+p collisions in PHENIX}

\author{Rapha\"el Granier de Cassagnac, for the PHENIX collaboration\footnote[3]{For the full PHENIX author list and acknowledgments, see Appendix "Collaborations" of this volume.}}

\address{Laboratoire Leprince-Ringuet, \'Ecole polytechnique/IN2P3, Palaiseau, 91128 France}

\ead{raphael@in2p3.fr}

\begin{abstract}
PHENIX has measured $J/\psi$ production at backward, central and forward rapidities both in p+p and d+Au collisions at $\sqrt{s_{NN}}=200$~GeV during the third run of the RHIC collider. From the p+p collisions, we measure the total J/$\psi$ cross section. From the d+Au collisions, we compute the nuclear modification factor versus transverse momentum, rapidity and centrality. All results presented here are preliminary.
\end{abstract}




\section{Introduction}

In nuclear collisions at 200~GeV, $J/\psi$s are mainly produced through gluon fusion and thus probe the gluon structure function and its modification in nuclei. 
$J/\psi$s are also predicted to be very sensitive to the creation of a quark-gluon plasma and are suppressed in Pb+Pb collisions at lower energy~\cite{NA50}. 
Measuring their yield in d+Au collisions is crucial to understand cold nuclear matter effects such as gluon shadowing, transverse momentum broadening, parton energy loss or absorption, as well as to set the baseline for future Au+Au measurements.

The PHENIX experiment is able to measure $J/\psi$s through their dilepton decay in four spectrometers: two central arms covering the mid-rapidity region of $|\eta|<0.35$ and twice $\pi/2$ in azimuth and two forward muon arms covering the full azimuth and $1.2 < |\eta| < 2.4$ in pseudorapidity. 
Electrons are identified in the central arms by their {\v C}erenkov rings and by matching the momentum of charged particles reconstructed in drift chambers with the energy deposited in an electromagnetic calorimeter. 
Muons are selected by an absorber and identified by the depth they reach in a succession of proportional counters staggered with steel walls.
The vertex and the event centrality are measured by beam-beam counters lying at $3 < |\eta| < 3.9$.

In this presentation, we report on $J/\psi$ measurements performed during the third RHIC run, with 2.74~nb$^{-1}$ d+Au and 350~nb$^{-1}$ p+p collisions. For this analysis, we use about 300 $J/\psi$ from d+Au and 100 from p+p observed in the central arms, and 1400 and 420 in the muon arms. 
Previous data samples have already been published by PHENIX and the analysis presented here is very similar to the one presented in \cite{PHENIX-Jpsipp} (150~nb$^{-1}$ p+p) and \cite{PHENIX-JpsiAuAu} (24~$\mu$b$^{-1}$ Au+Au). All results are preliminary.

\section{$J/\psi$ cross section in proton+proton}

\begin{figure}
 \begin{minipage}[b]{.46\linewidth}
  \flushright\epsfig{figure=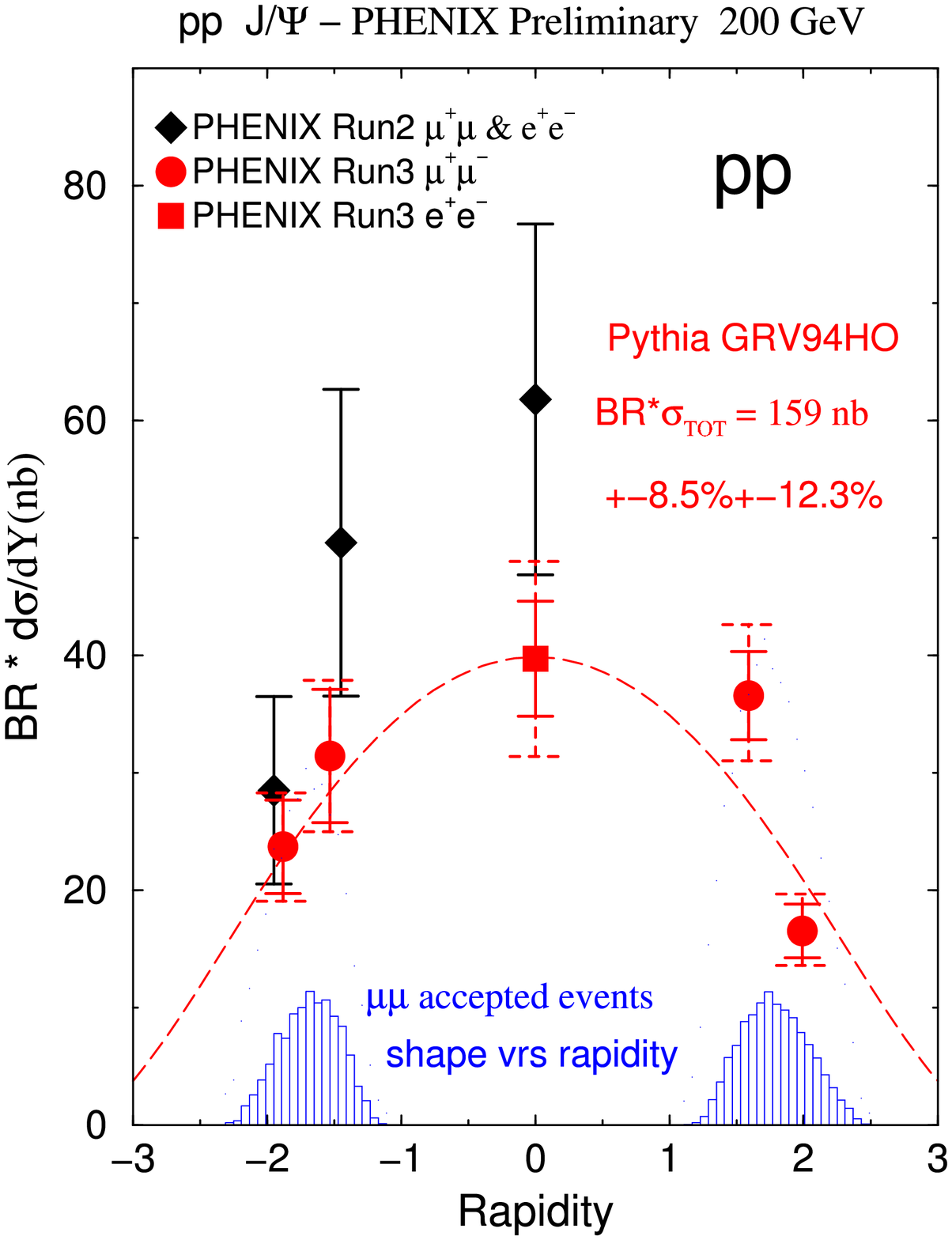,width=.75\linewidth}
  \vspace*{2ex}\caption{$J/\psi$ differential cross section, multiplied by the dilepton branching ratio, versus rapidity.\label{fig:dsigdy_pp}}
 \end{minipage} \hfill
 \begin{minipage}[b]{.46\linewidth}
  \flushright\epsfig{figure=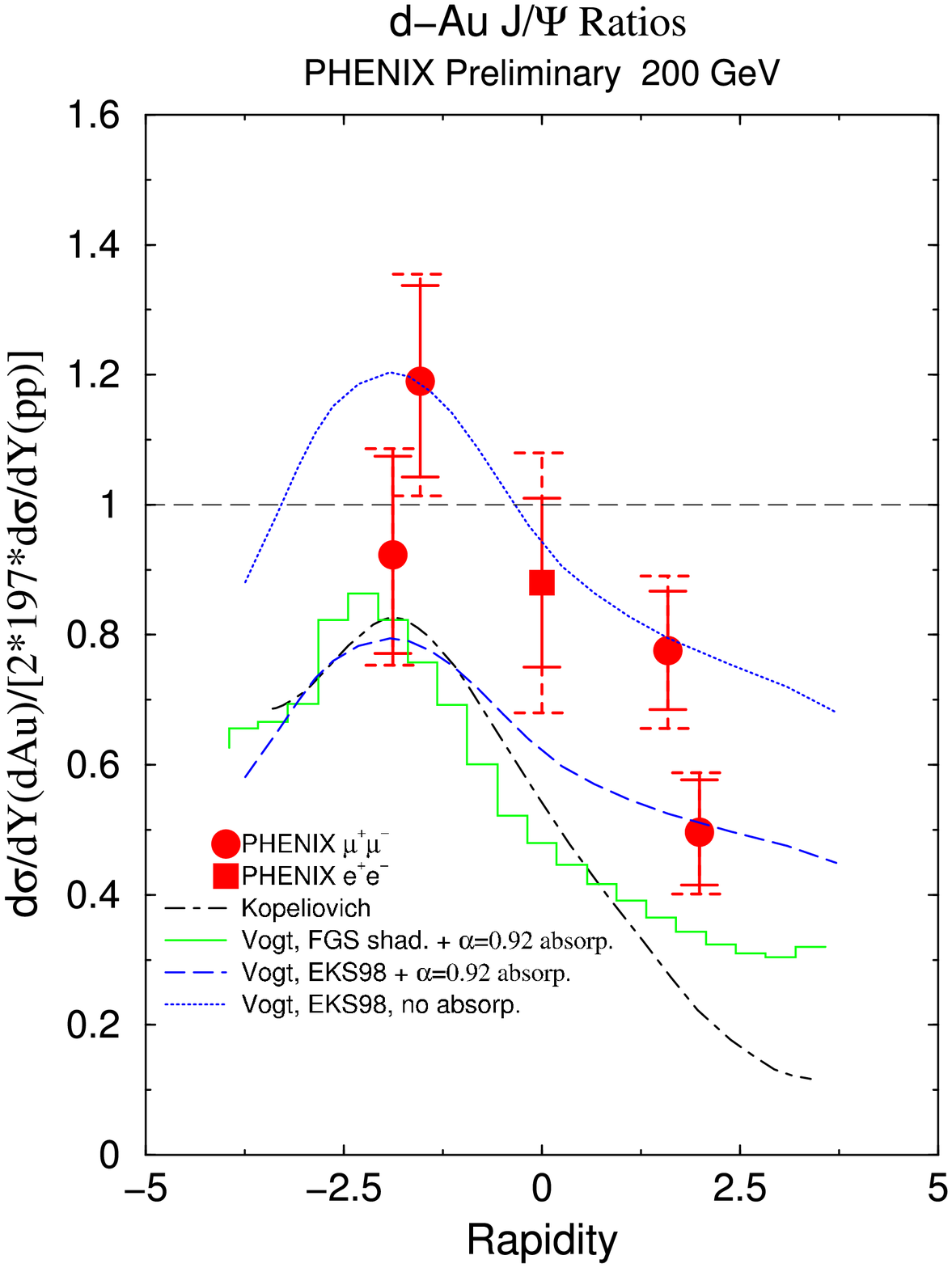,width=.75\linewidth}
  \vspace*{2ex}\caption{Ratio between d+Au and p+p $J/\psi$ differential cross sections, divided by $2ccccc\times 197$, versus rapidity. \label{fig:dau_y}}
 \end{minipage}
\end{figure}

On figure \ref{fig:dsigdy_pp}, we show the measured differential cross section versus rapidity, multiplied by the dilepton branching ratio. The dielectron measurement is shown as a single point at mid-rapidity. The data from each dimuon spectrometer has been split in two rapidity bins. Solid error bars represent statistical and point to point systematic uncertainties. The dashed error bars stand for the systematic uncertainties common to one spectrometer. An additional 12.3\% global error bar is not displayed.

A fit to a shape generated with PYTHIA using the GRV94HO parton distribution is performed and gives a total cross section, multiplied by the dilepton branching ratio of 5.9\%, equal to:
\begin{equation}
{\rm BR} \times \sigma_{pp}^{J/\psi} = 159 \; {\rm nb} \pm 8.5\% ({\rm fit}) \pm 12.3\% ({\rm abs})
\end{equation}
where the first uncertainty comes from the fit and thus includes both the statistical and point-to-point systematics. The second uncertainty accounts for absolute systematic errors.  Variations in the parton distribution functions used to determine the shape was demonstrated to be less than 3\%~\cite{PHENIX-Jpsipp} and was neglected here. 

\section{$J/\psi$ production in deuteron+gold}

In d+Au collisions at 200 GeV, $J/\psi$s in our three rapidity ranges probe the following momentum fractions $x$ of gluons in the gold nucleus (neglecting the emitted gluon): 0.05~to~0.14~(backward, negative rapidity, gold-going side), 0.011~to~0.022~(midrapidity) and 0.0014~to~0.0047~(forward, positive rapidity, deuteron-going side).

\subsection{Rapidity dependence}

The ratio between the $J/\psi$ yields observed in d+Au collisions and p+p collisions, divided by $2\times197$ is shown on figure~\ref{fig:dau_y}. Solid error bars represent statistical and point to point systematic uncertainties. The dashed error bars stand for the systematic uncertainties common to one spectrometer. An additional 13.4\% global error bar is not displayed. 

While this ratio is close to unity at backward rapidity, it is significantly lower at forward rapidity, where parton distribution are expected to be shadowed in a heavy nucleus. Theoretical predictions from Vogt~\cite{THEORY-Vogt} and Kopeliovich~\cite{THEORY-Kopeliovich} are displayed on the figure for comparison.

\subsection{Transverse momentum dependence}

The $J/\psi$ $p_T$ distributions $(d\sigma/d p_T)/(2\pi p_T)$ have been fitted to the traditional $(A(1+(p_T/B)^2)^{-6}$ function. The average $p^2_T$ resulting from the fit is $4.47 \pm 0.25$ and $3.99 \pm 0.25$ for d+Au collisions at backward and forward rapidity, respectively, to be compared with  $2.70 \pm 0.24$ for p+p collisions.

Figure~\ref{fig:alpha_pt} shows the $\alpha$ parameter defined as $\sigma_{dA} = \sigma_{pp} \times (2A)^\alpha$. Solid error bars represent statistical and point to point systematic uncertainties. An additional 2.2\% global error bar is not displayed. Results from the $\sqrt{s_{NN}}=38$~GeV lower energy experiment E866/NuSea~\cite{E866} are displayed for comparison. The amounts of $p_T$ broadening are comparable at both energies.

\subsection{Centrality dependence}

\begin{figure}
\begin{minipage}{7.5cm}
\centering\epsfig{file=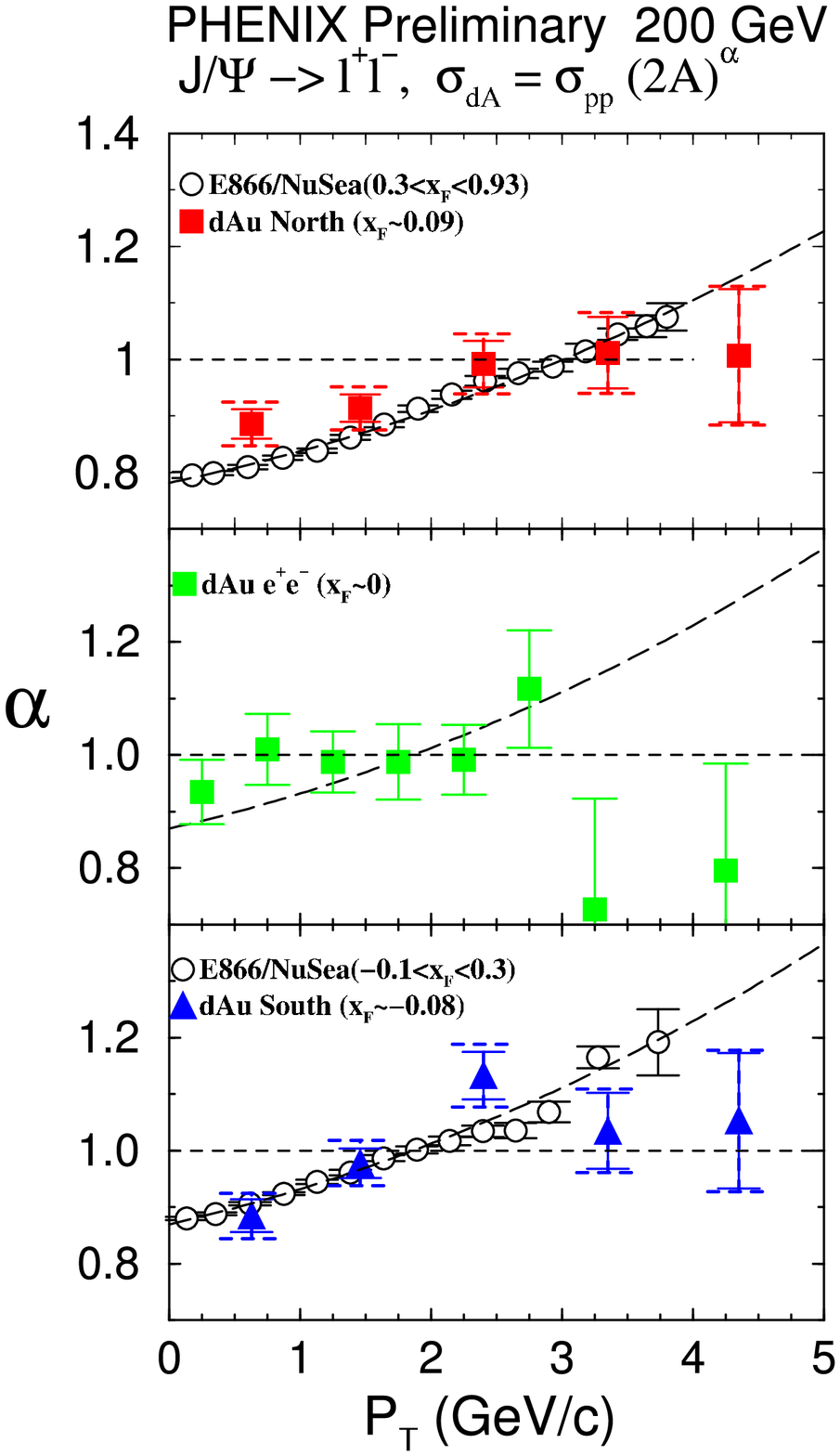,width=3.8cm}
\end{minipage}
\begin{minipage}{7.5cm}
\epsfig{file=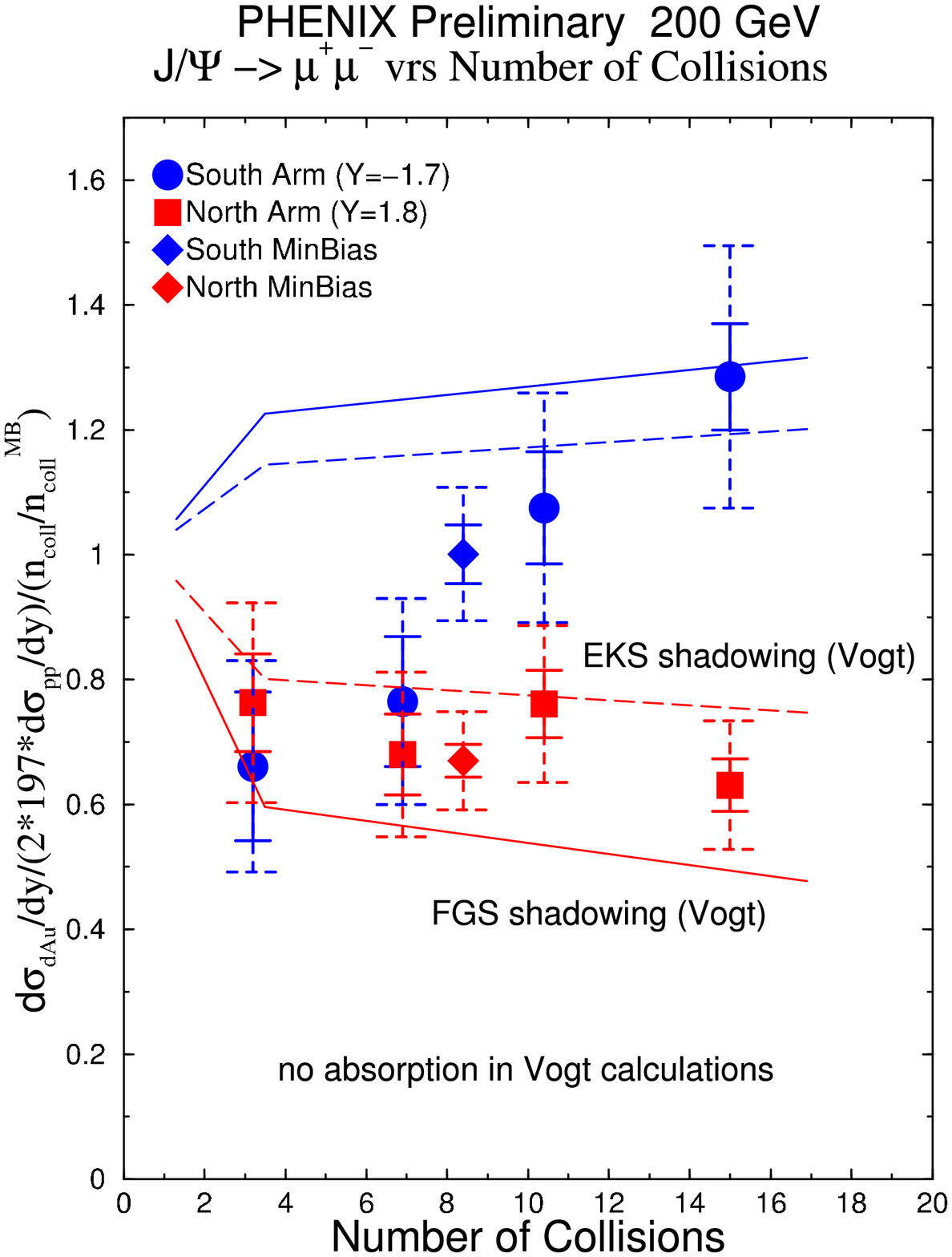,width=6.5cm}
\end{minipage}
\vspace{2ex}
\begin{minipage}{7.5cm}
\caption{\label{fig:alpha_pt} $\alpha$ parameter versus $p_T$ compared to lower energy measurements.}
\end{minipage}
\begin{minipage}{7.5cm}
\caption{\label{fig:rda_ncoll} Nuclear modification factor versus number of collisions.}
\end{minipage}
\end{figure}

For d+Au collisions, we measure centrality by counting the charge deposited in the beam-beam counter facing the gold beam. Centrality is then related to the average number of nucleon+nucleon collisions $<N_{coll}>$ through a Glauber computation. The ratio $R_{dA}$ of $J/\psi$ yields in d+Au and p+p normalized by the $<N_{coll}>$ is shown on figure~\ref{fig:rda_ncoll}, for four centrality classes corresponding to $<N_{coll}> = 3.2 \pm 0.3$, $6.9 \pm 0.6$, $10.4 \pm 0.7$ and $15.0 \pm 1.0$, and for minimum bias collisions corresponding to an average value of $N_{coll}$ equal to $8.4 \pm 0.7$. Solid error bars represent statistical and point to point systematics and the dashed error bars are common to one spectrometer. An additional 13.4\% global error bar is not displayed.

At forward rapidity (low $x$ values in the gold nucleus), no strong dependence is observed, while a strong enhancement  from peripheral to central collisions is observed at backward rapidity. The theoretical curves on figure~\ref{fig:rda_ncoll} stand for corresponding amounts of inhomogeneous shadowing and antishadowing at these rapidities from~\cite{THEORY-Vogt}. The predictions are qualitatively consistent with the data at forward rapidity, while they do not show the observed steep rising shape at backward rapidity. For now, we give no interpretation to this unexpected behaviour, but speculate that it is related to the fact that these $J/\psi$s are closer to the gold nucleus rest frame. 

\section{Conclusions}

During run 3, the PHENIX experiment was able to see nuclear effects affecting $J/\psi$ production in d+Au collisions. The rapidity dependence is consistent with shadowing of gluon distribution functions and the $p_T$ differential cross section shapes exhibit broadening. However, given the limited statistics, it is difficult to quantify the level of $J/\psi$ absorption or to distinguish between various theoretical models of shadowing. A surprising rising shape was seen in the centrality dependence at backward rapidity and waits for interpretation.

\end{document}